\documentclass[12pt,twoside]{article}
\usepackage[english]{babel}
\usepackage{graphicx,rotating,epsfig}
\usepackage{a4p}

\def\ra{\rightarrow}
\def\GeV  {\ensuremath{\mathrm{ Ge\kern -0.1em V } }}
\def\GeVc2{\ensuremath{\mathrm{ Ge\kern -0.1em V }\kern -0.2em /c^2 }}
\def\MeVc2{\ensuremath{\mathrm{ Me\kern -0.1em V }\kern -0.2em /c^2 }}
\newcommand{\MT}{\ensuremath{M_{\mathrm{t}}}}
\newcommand{\MTll}{\ensuremath{\MT^{\mathrm{di-l}}}}
\newcommand{\MTlj}{\ensuremath{\MT^{\mathrm{l+j}}}}
\newcommand{\MTjj}{\ensuremath{\MT^{\mathrm{all-j}}}}

\newcommand{\pb}{\ensuremath{\mathrm{pb}^{-1}}}
\newcommand{\ttbar}{\ensuremath{t\overline{t}}}
\newcommand{\ljt}{\ensuremath{\ell\nu q q^{\prime} b \overline{b}}}
\newcommand{\had}{\ensuremath{q q^{\prime} b q q^{\prime} \overline{b}}}
\newcommand{\dil}{\ensuremath{\ell^{+}\nu b\ell^{-}\overline{\nu}\overline{b}}}
\newcommand{\ttljt}{\ensuremath{\ttbar\ra\ljt}}
\newcommand{\ttdil}{\ensuremath{\ttbar\ra\dil}}
\newcommand{\tthad}{\ensuremath{\ttbar\ra\had}}
\newcommand{\RunI}{\hbox{Run-I}}
\newcommand{\RunII}{\hbox{Run-II}}

\begin{document}

\begin{center}
  {\LARGE FERMI NATIONAL ACCELERATOR LABORATORY}
\end{center}

\begin{flushright}
       FERMILAB-TM-2351-E \\  
       TEVEWWG/top 2006/02 \\
       CDF Note 8231 \\
       D\O\ Note 5098 \\
       hep-ex/0604053 \\
       \vspace*{0.05in}

       {27-April-2006}
\end{flushright}

\vskip 1cm

\begin{center}
  {\LARGE\bf 
    Combination of CDF and D\O\ Results \\
    on the Mass of the Top Quark
  }
  \vfill
  {\Large
    The Tevatron Electroweak Working Group\footnote{WWW access at 
    {\tt http://tevewwg.fnal.gov}\\
    The members of the TEVEWWG who contributed significantly to the
    analysis described in this note are: \\
    E.~Brubaker (brubakee@fnal.gov),        
    F.~Canelli (canelli@fnal.gov),          
    R.~Demina (demina@fnal.gov),            
    R.~Erbacher (robine@fnal.gov),          
    I.~Fleck (fleck@fnal.gov),              
    D.~Glenzinski (douglasg@fnal.gov),      
    G.~Gutierrez (gaston@fnal.gov),         
    M.~W.~Gr\"unewald (mwg@fnal.gov),       
    A.~Juste (juste@fnal.gov),              
    T.~Maruyama (maruyama@fnal.gov),        
    A.~Quadt (quadt@fnal.gov),              
    E.~Thomson (thomsone@fnal.gov),         
    C.~Tully (tully@fnal.gov),              
    E.~W.~Varnes (varnes@fnal.gov),         
    D.~O.~Whiteson (danielw@fnal.gov),      
    U.K.~Yang (ukyang@fnal.gov).            
    } \\ for the CDF and D\O\ Collaborations 
  }
\end{center}
\vfill
\begin{abstract}
\noindent
  We summarize the published top-quark mass measurements from the CDF
  and D\O\ experiments at Fermilab.  We combine published \RunI\
  (1992-1996) measurements with the most recent published \RunII\
  (2001-present) measurements using up to $340~\pb$ of data.  Taking
  correlated uncertainties properly into account the resulting mass of
  the top quark is
  $\MT=174.2\pm2.0\mathrm{(stat)}\pm2.6\mathrm{(syst)}~\GeVc2$, which
  corresponds to a total uncertainty of $3.3~\GeVc2$, i.e. 1.9\%
  precision.\\ 
  Since this combination
  uses only a subset of the available analyses and data sets, it does
  not supersede our latest world average combination of
  $\MT=172.5\pm1.3\mathrm{(stat)}\pm1.9\mathrm{(syst)}~\GeVc2$, which
  is based on the latest published and preliminary results.
%
\end{abstract}

\vfill



\section{Introduction}
\label{sec:intro}

The experiments CDF and D\O, taking data at the Tevatron
proton-antiproton collider located at the Fermi National Accelerator
Laboratory, have made several direct experimental measurements of the
top-quark pole mass, \MT.  The pioneering measurements were based on about
$100~\pb$ of \RunI\ (1992-1996) data~\cite{Mtop1-CDF-di-l-PRLa, 
  Mtop1-CDF-di-l-PRLb,
  Mtop1-CDF-di-l-PRLb-E, Mtop1-D0-di-l-PRL, Mtop1-D0-di-l-PRD,
  Mtop1-CDF-l+j-PRL, Mtop1-CDF-l+j-PRD, Mtop1-D0-l+j-old-PRL,
  Mtop1-D0-l+j-old-PRD, Mtop1-D0-l+j-new1, Mtop1-CDF-all-j-PRL,
  Mtop1-D0-all-j-PRL} 
and include results from the \tthad\ (all-j), the \ttljt\ (l+j), and the 
\ttdil\ (di-l) decay channels\footnote{Here $\ell=e$ or $\mu$.  Decay 
channels with explicit tau lepton identification are presently under
study and are not yet used for measurements of the top-quark mass.}.
Results using approximately $340~\pb$ of \RunII\ (2001-present) data
have been recently published in the l+j and di-l
channels~\cite{Mtop2-CDF-l+j-350PRL, Mtop2-CDF-l+j-350PRD,
Mtop2CDF-di-l-350PRL}. More preliminary analyses have been performed
in the l+j and di-l channels using $370-750~\pb$ of data and improved
analysis techniques~\cite{Mtop2-CDF-l+j-new,
Mtop2-CDF-di-l-new,cdf_hepex0512009,cdf_hepex0602008,cdf_note8151,cdf_note8133,cdf_note7718,
Mtop2-D0-l+j-new,Mtop2-D0-di-l-new,d0_note4574,d0_note4728,d0_note5047}.
\vspace*{0.10in}

This note reports the average top quark mass obtained by combining
five published \RunI\ measurements~\cite{Mtop1-CDF-di-l-PRLb,
Mtop1-CDF-di-l-PRLb-E, 
Mtop1-D0-di-l-PRL, 
Mtop1-CDF-l+j-PRD,
Mtop1-D0-l+j-new1, Mtop1-CDF-all-j-PRL} with the most recent published
\RunII\ measurements from
CDF~\cite{Mtop2-CDF-l+j-350PRL, Mtop2-CDF-l+j-350PRD,
Mtop2CDF-di-l-350PRL}.\footnote{D\O\ results
in the l+j and di-l channels using Run II data are at present close to
submission but not yet published and are therefore not included in
this combination.}
%
%
Since this combination only uses a subset of
the available analyses and data sets, it does not supersede our latest
world average combination of
$\MT=172.5\pm1.3\mathrm{(stat)}\pm1.9\mathrm{(syst)}~\GeVc2$~\cite{Mtop-tevewwgWin06}.
The combination takes into account the statistical and systematic
uncertainties and their correlations using the method of
references~\cite{Lyons:1988, Valassi:2003}. The most precise
individual measurements of $\MT$ in this combination are the
measurements in the l+j channel from Run I and Run II.  These are
$173.5^{+3.9}_{-3.8}\;\rm GeV/c^2$ (CDF II,
\cite{Mtop2-CDF-l+j-350PRL}) and $180.1
\pm 5.3\;\rm GeV/c^2$ (D\O\ I, \cite{Mtop1-D0-l+j-new1}). These have
weights in the new $\MT$ combination of 57\% and 26\%, respectively.
\vspace*{0.10in}

The input measurements and error categories used in the combination are 
detailed in Section~\ref{sec:inputs} and~\ref{sec:errors}, respectively. 
The correlations used in the combination are discussed in 
Section~\ref{sec:corltns} and the resulting world average top-quark mass 
is given in Section~\ref{sec:results}.  A summary and outlook are presented
in Section~\ref{sec:summary}.
 
\section{Input Measurements}
\label{sec:inputs}

For this combination seven measurements of \MT\ are used, five published
\RunI\ results and two published \RunII\ results. The most precise 
result in each channel is considered for each experiment. In general, 
the \RunI\ measurements all have relatively large statistical
uncertainties and their systematic uncertainty is dominated by the
total jet energy scale (JES) uncertainty.  In \RunII\ both CDF and
D\O\ take advantage of the larger
\ttbar\ samples available and employ new analysis techniques to reduce
both these uncertainties.  In particular the JES is constrained
using an in-situ calibration based on the invariant mass of $W\ra qq^{\prime}$ 
decays in the l+j channel.  
The \RunII\ CDF analysis in the l+j channel 
constrains the 
JES using the in-situ $W\ra qq^{\prime}$ decays.  The JES is also determined 
using ``external'' calibration samples as was done for the \RunI\ measurements.
This external JES is applied as an additional constraint in this CDF analysis 
to improve further the total JES uncertainty. Small residual JES uncertainties 
arising from $\eta-$ and $p_{T}$-dependencies and the modeling of $b$-jets 
are included in 
separate error categories.   The \RunII\ CDF measurement in the di-l channel 
uses only the externally determined JES, some parts of which are correlated 
with the externally determined \RunI\ JES as noted below.
\vspace*{0.10in}

The CDF \RunII\ measurement in the l+j channel requires special treatment 
in order to account more accurately for the correlations of the JES 
uncertainties since the fit uses information from both the in-situ and 
external JES calibrations.  In the 
combination we treat this one measurement as two separate inputs - one
which includes only the in-situ JES calibration, (l+j)$_i$, and a second
which includes only the JES as determined from the external calibration
samples, (l+j)$_e$.  We correlate the JES related error categories as
described below while taking the rest of the error categories to be 100\%
correlated between these two inputs.  The combination of just these two 
inputs using these correlations yields the identical central value,
statistical, JES, and total systematic uncertainty as the measurement 
reported in reference~\cite{Mtop2-CDF-l+j-new}.  The correlations between these
two inputs and the rest of the inputs are as described in Section~\ref{sec:corltns}.
\vspace*{0.10in}

The inputs used in the combination are summarized in Table~\ref{tab:inputs} 
with their uncertainties sub-divided into the categories described in the
next Section.

\begin{table}[t]
\begin{center}
\renewcommand{\arraystretch}{1.30}
\begin{tabular}{|l||rrr|rr||rrr|}
\hline       
       & \multicolumn{5}{|c||}{{\RunI} published} & \multicolumn{3}{|c|}{{\RunII} published} \\ \cline{2-9}
       & \multicolumn{3}{|c|}{ CDF } & \multicolumn{2}{|c||}{ D\O\ }
       & \multicolumn{3}{|c|}{ CDF } \\
       & all-j &   l+j &  di-l &   l+j &  di-l & (l+j)$_i$ & (l+j)$_e$ & di-l  \\
\hline                         
Lumi (\pb) & 110 & 105 & 110 & 125 & 125 & 320 & 320 & 340  \\\hline
Result & 186.0 & 176.1 & 167.4 & 180.1 & 168.4 & 173.5 & 173.5 & 165.2  \\
\hline                         
\hline                         
iJES   &   0.0 &   0.0 &   0.0 &   0.0 &   0.0 &   4.2 &   0.0 &   0.0  \\
bJES   &   0.6 &   0.6 &   0.8 &   0.7 &   0.7 &   0.6 &   0.6 &   0.5  \\
cJES   &   3.0 &   2.7 &   2.6 &   2.0 &   2.0 &   0.0 &   2.0 &   2.2  \\
dJES   &   0.3 &   0.7 &   0.6 &   0.0 &   0.0 &   0.0 &   0.6 &   0.8  \\
rJES   &   4.0 &   3.4 &   2.7 &   2.5 &   1.1 &   0.0 &   2.2 &   1.1  \\
Signal &   1.8 &   2.6 &   2.8 &   1.1 &   1.8 &   0.8 &   0.8 &   1.3  \\
BG     &   1.7 &   1.3 &   0.3 &   1.0 &   1.1 &   0.5 &   0.5 &   0.8  \\
Fit    &   0.6 &   0.0 &   0.7 &   0.6 &   1.1 &   0.6 &   0.6 &   1.3  \\
MC     &   0.8 &   0.1 &   0.6 &   0.0 &   0.0 &   0.2 &   0.2 &   0.8  \\
UN/MI  &   0.0 &   0.0 &   0.0 &   1.3 &   1.3 &   0.0 &   0.0 &   0.0  \\
\hline                         
Syst.  &   5.7 &   5.3 &   4.8 &   3.9 &   3.6 &   4.4 &   3.3 &   3.4  \\
Stat.  &  10.0 &   5.1 &  10.3 &   3.6 &  12.3 &   2.7 &   2.7 &   6.1  \\
\hline                         
\hline                         
Total  &  11.5 &   7.3 &  11.4 &   5.3 &  12.8 &   5.2 &   4.3 &   7.0  \\ \hline\hline
\end{tabular}
\end{center}
\caption[Input measurements]{Summary of the measurements used to determine this
  average $\MT$ using only published $\MT$ results.  As described in
  the text, the CDF {\RunII} measurement in the lepton+jets channel is
  treated as two inputs in order to more accurately account for the
  correlations in the jet energy scale uncertainties.  All numbers are
  in $\GeVc2$.  The error categories and their correlations are
  described in the text.  The total systematic uncertainty and the
  total uncertainty are obtained by adding the relevant contributions
  in quadrature.}
\label{tab:inputs}
\end{table}


\section{Error Categories}
\label{sec:errors}

We employ the same error categories as used for the previous world
average~\cite{Mtop-tevewwgSum05}.  They have evolved to include a detailed
breakdown of the various sources of uncertainty and aim to
lump together sources of systematic uncertainty that share the same or
similar origin.  For example, the ``Signal'' category discussed below
includes the uncertainties from ISR, FSR, and PDF - all of which affect
the modeling of the \ttbar\ signal.  Additional categories have been
added in order to accommodate specific types of correlations.  For example,
the jet energy scale (JES) uncertainty is sub-divided into several
components in order to more accurately accommodate our best estimate of
the relevant correlations.  Each error category is discussed below.
\vspace*{0.10in}

\begin{description}
  \item[Statistical:] The statistical uncertainty associated with the
    \MT\ determination.
  \item[iJES:] That part of the JES uncertainty which originates from 
    in-situ calibration procedures and is uncorrelated among the
    measurements.  In the combination reported here it corresponds to  
    the statistical uncertainty associated with the JES determination 
    using the $W\ra qq^{\prime}$ invariant mass in the CDF \RunII\ 
    l+j measurement.  Residual JES uncertainties, which arise from effects 
    not considered in the in-situ calibration, are included in other 
    categories.
%
%
  \item[bJES:] That part of the JES uncertainty which originates from
    uncertainties specific to the modeling of $b$-jets and which is correlated
    across all measurements.  For both CDF and D\O\ this includes uncertainties 
    arising from 
    variations in the semi-leptonic branching fraction, $b$-fragmentation 
    modeling, and differences in the color flow between $b$-jets and light-quark
    jets.  These were determined from \RunII\ studies but back-propagated
    to the \RunI\ measurements, whose rJES uncertainties (see below) were 
    then corrected in order to keep the total JES uncertainty constant.
  \item[cJES:] That part of the JES uncertainty which originates from
    modeling uncertainties correlated across all measurements.  Specifically
    it includes the modeling uncertainties associated with light-quark 
    fragmentation and out-of-cone corrections.
  \item[dJES:] That part of the JES uncertainty which originates from
    limitations in the calibration data samples used and which is 
    correlated between measurements within the same data-taking period
    (i.e. Run~I or Run~II) but not between experiments.  For CDF this
    corresponds to uncertainties associated with the $\eta$-dependent JES 
    corrections which are estimated using di-jet data events.  
%
  \item[rJES:] The remaining part of the JES uncertainty which is 
    correlated between all measurements of the same experiment 
    independent of data-taking period, but is uncorrelated between
    experiments.  This is dominated by uncertainties in the calorimeter
    response to light-quark jets.  For CDF this also includes small 
    uncertainties associated with the multiple interaction and underlying 
    event corrections.
  \item[Signal:] The systematic uncertainty arising from uncertainties
    in the modeling of the \ttbar\ signal which is correlated across all
    measurements.  This includes uncertainties from variations in the ISR,
    FSR, and PDF descriptions used to generate the \ttbar\ Monte Carlo samples
    that calibrate each method.  It also includes small uncertainties 
    associated with biases associated with the identification of $b$-jets.
  \item[Background:]  The systematic uncertainty arising from uncertainties
    in modeling the dominant background sources and correlated across
    all measurements in the same channel.  These
    include uncertainties on the background composition and shape.  In
    particular uncertainties associated with the modeling of the QCD
    multi-jet background (all-j and l+j), uncertainties associated with the
    modeling of the Drell-Yan background (di-l), and uncertainties associated 
    with variations of the fragmentation scale used to model W+jets 
    background (all channels) are included.
  \item[Fit:] The systematic uncertainty arising from any source specific
    to a particular fit method, including the finite Monte Carlo statistics 
    available to calibrate each method.
  \item[Monte Carlo:] The systematic uncertainty associated with variations
    of the physics model used to calibrate the fit methods and correlated
    across all measurements.  For CDF it includes variations observed when 
    substituting PYTHIA~\cite{PYTHIA4,PYTHIA5,PYTHIA6} (Run~I and Run~II) 
    or ISAJET~\cite{ISAJET} (Run~I) for HERWIG~\cite{HERWIG5,HERWIG6} when 
    modeling the \ttbar\ signal.  Similar
    variations are included for the D\O\ \RunI\ measurements.  
%
  \item[UN/MI:] This is specific to D\O\ and includes the uncertainty
    arising from uranium noise in the D\O\ calorimeter and from the
    multiple interaction corrections to the JES.  
%
\end{description}
These categories represent the current preliminary understanding of the
various sources of uncertainty and their correlations.  We expect these to 
evolve as we continue to probe each method's sensitivity to the various 
systematic sources with ever improving precision.  Variations in the assignment
of uncertainties to the error categories, in the back-propagation of the bJES
uncertainties to \RunI\ measurements, in the approximations made to
symmetrize the uncertainties used in the combination, and in the assumed 
magnitude of the correlations all negligibly effect ($\ll 100\;\MeVc2$) the 
combined \MT\ and total uncertainty.

\section{Correlations}
\label{sec:corltns}

The following correlations are used when making the combination:
\begin{itemize}
  \item The uncertainties in the Statistical, Fit, and iJES
    categories are taken to be uncorrelated among the measurements.
  \item The uncertainties in the 
    dJES category
    is taken
    to be 100\% correlated among all \RunI\ and all \RunII\ measurements 
    on the same experiment, but uncorrelated between Run~I and Run~II
    and uncorrelated between the experiments.
  \item The uncertainties in the rJES and UN/MI categories are taken
    to be 100\% correlated among all measurements on the same experiment.
  \item The uncertainties in the Background category are taken to be
    100\% correlated among all measurements in the same channel.
  \item The uncertainties in the bJES, cJES, Signal, and Generator
    categories are taken to be 100\% correlated among all measurements.
\end{itemize}
Using the inputs from Table~\ref{tab:inputs} and the correlations specified
here, the resulting matrix of total correlation co-efficients is given in
Table~\ref{tab:coeff}.

\begin{table}[t]
\begin{center}
\renewcommand{\arraystretch}{1.30}
\begin{tabular}{|ll||rrr|rr||rrr|}
\hline       
   &   & \multicolumn{5}{|c||}{{\RunI} published} & \multicolumn{3}{|c|}{{\RunII} published} \\ \cline{3-10}
   &   & \multicolumn{3}{|c|}{ CDF } & \multicolumn{2}{|c||}{ D\O\ }
       & \multicolumn{3}{|c|}{ CDF }  \\
   &  &   l+j   &  di-l & all-j &   l+j &  di-l & (l+j)$_i$ & (l+j)$_e$ & di-l  \\
\hline
\hline
CDF-I & l+j     &   1.00&       &       &       &       &       &       &       \\
CDF-I & di-l    &   0.29&   1.00&       &       &       &       &       &       \\
CDF-I & all-j   &   0.32&   0.19&   1.00&       &       &       &       &       \\
\hline
D\O-I & l+j     &   0.26&   0.15&   0.14&   1.00&       &       &       &       \\
D\O-I & di-l    &   0.11&   0.08&   0.07&   0.16&   1.00&       &       &       \\
\hline
\hline
CDF-II & (l+j)$_i$& 0.08&   0.05&   0.03&   0.07&   0.03&   1.00&       &       \\
CDF-II & (l+j)$_e$& 0.51&   0.29&   0.34&   0.26&   0.11&   0.41&   1.00&       \\
CDF-II & di-l   &   0.26&   0.17&   0.18&   0.17&   0.09&   0.04&   0.30&  1.00 \\
\hline
\end{tabular}
\end{center}
\caption[Global correlations between input measurements]{The resulting
  matrix of total correlation coefficients used in the top-quark mass combination
  reported here.}
\label{tab:coeff}
\end{table}

The measurements are combined using a program implementing a numerical
$\chi^2$ minimization as well as the analytic BLUE
method~\cite{Lyons:1988, Valassi:2003}. The two methods used are
mathematically equivalent, and are also equivalent to the method used
in an older combination~\cite{TM-2084}, and give identical results for
the combination. In addition, the BLUE method yields the decomposition
of the error on the average in terms of the error categories specified
for the input measurements~\cite{Valassi:2003}.

\section{Results}
\label{sec:results}

\begin{figure}[p]
\begin{center}
\includegraphics[width=0.8\textwidth]{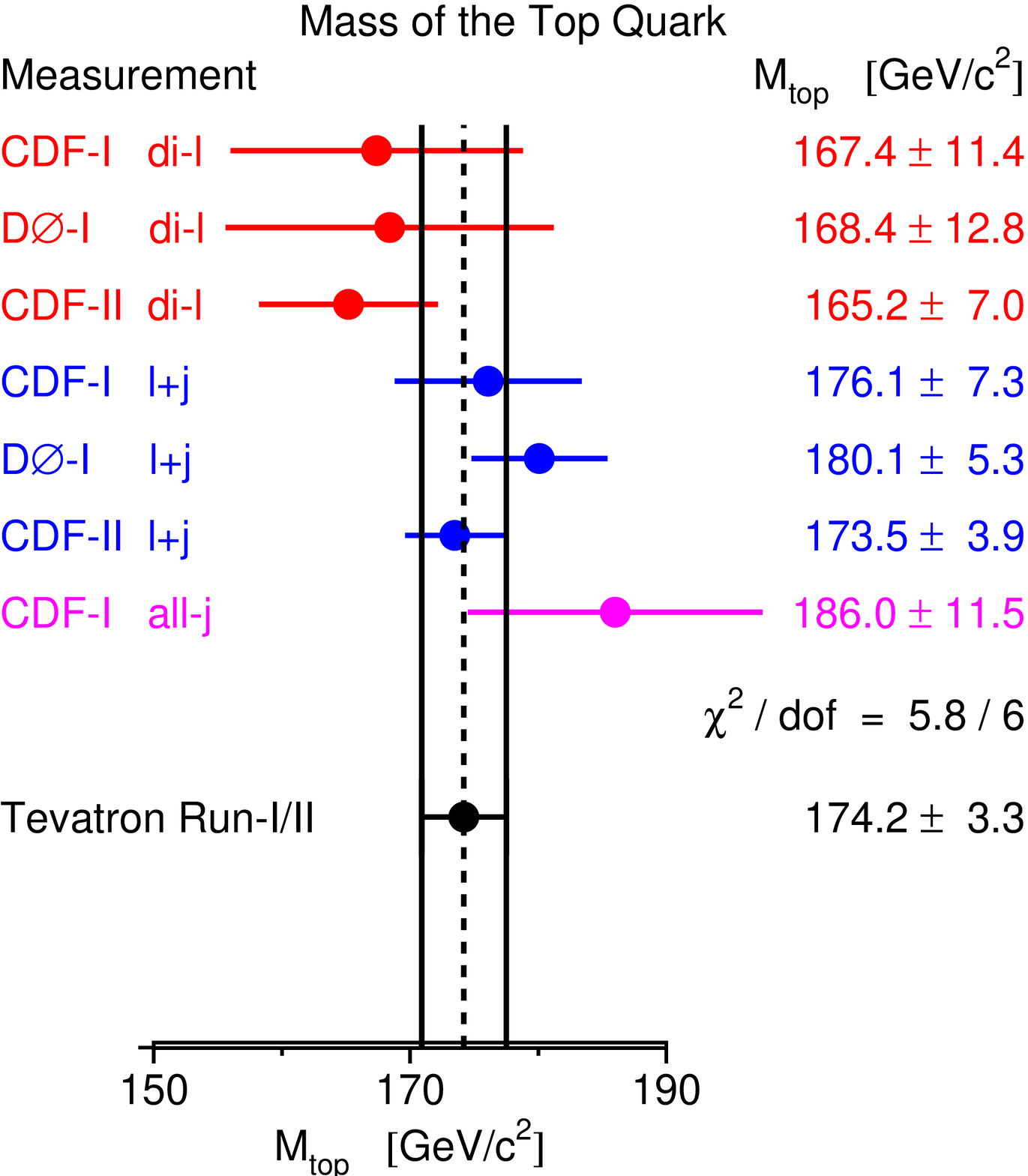}
\end{center}
\caption[Summary plot for the world average top-quark mass]
  {A summary of the input measurements and the resulting 
   combined top-quark mass.}
\label{fig:summary} 
\end{figure}

The combined value for the top-quark mass is:
\begin{eqnarray}
  \MT & = & 174.2 \pm 3.3~\GeVc2\,,
\end{eqnarray}
with a $\chi^2$ of 5.8 for 6 degrees of freedom, which corresponds to
a probability of 45\% indicating good agreement among all the input
measurements.  The total uncertainty can be sub-divided into the 
contributions from the various error categories as: Statistical ($\pm2.0$),
total JES ($\pm2.3$), Signal ($\pm1.0$), Background ($\pm0.6$), Fit
($\pm0.4$), Monte Carlo ($\pm0.2$), and UN/MI ($\pm0.4$), for a total
Systematic ($\pm2.6$), where all numbers are in units of \GeVc2.
The pull and weight for each of the inputs are listed in Table~\ref{tab:stat}.
The input measurements and the resulting combined to-quark mass
are summarized in Figure~\ref{fig:summary}.
\vspace*{0.10in}

In this combination, for the CDF \RunII\ l+j measurement using 320~\pb
of data, the in-situ and external JES calibrated inputs each carry
approximately the same weight. In the latest world average combination
reported Reference~\cite{Mtop-tevewwgWin06}, the weight of the CDF
\RunII\ l+j input using the in-situ JES calibration carries three
times the weight of its counterpart using the external JES
calibration.  This trend is expected to continue with more data since
the in-situ JES uncertainty is expected to improve as the statistics
of the $W\ra qq^{\prime}$ sample increase with larger data sets.  In
contrast the uncertainty on the external JES calibration already has
large contributions from modeling uncertainties which may not be
reduced with larger data sets.
\vspace*{0.10in}


\begin{table}[t]
\begin{center}
\renewcommand{\arraystretch}{1.30}
\begin{tabular}{|l||rrr|rr||rrr|}
\hline       
       & \multicolumn{5}{|c||}{{\RunI} published} & \multicolumn{3}{|c|}{{\RunII} published} \\ \cline{2-9}
       & \multicolumn{3}{|c|}{ CDF } & \multicolumn{2}{|c||}{ D\O\ }
       & \multicolumn{3}{|c|}{ CDF } \\
       & l+j & di-l & all-j & l+j  & di-l  & (l+j)$_i$ & (l+j)$_e$ & di-l \\
\hline
\hline
Pull   & $+0.30$ & $-0.62$ & $+1.07$ & $+1.43$ &  $-0.46$ & $-0.17$ & $-0.24$ & $-1.45$ \\
Weight [\%]
       & $ 1.0$ & $ 1.1$ & $ 0.6$ & $26.2$ &  $ 2.8$ & $28.2$ & $28.5$ & $11.6$  \\
\hline
\end{tabular}
\end{center}
\caption[Pull and weight of each measurement]{The pull and weight for each of the
  inputs used to determine this average mass of the top quark.}
\label{tab:stat} 
\end{table} 

Although the $\chi^2$ from the combination of all measurements indicates
that there is good agreement among them, and no input has an anomalously
large pull, it is still interesting to also fit for the top-quark mass
in the all-j, l+j, and di-l channels separately.  We use the same methodology,
inputs, error categories, and correlations as described above, but fit for
the three physical observables, \MTjj, \MTlj, and \MTll.
The results of this combination are shown in Table~\ref{tab:three_observables}
and have $\chi^2$ of 1.4 for 5 degrees of freedom, which corresponds to a
probability of 93\%.
These results differ from a naive combination, where
only the measurements in a given channel contribute to the \MT\ 
determination in that channel, since the combination here fully accounts
for all correlations, including those which cross-correlate the different
channels. 

\begin{table}[t]
\begin{center}
\renewcommand{\arraystretch}{1.30}
\begin{tabular}{|l||c|rrr|}
\hline
Parameter & Value (\GeVc2) & \multicolumn{3}{|c|}{Correlations} \\
\hline
\hline
$\MTjj$ & $186.7\pm          11.0$ & 1.00 &      &      \\
$\MTlj$ & $175.6\pm\phantom{0}3.4$ & 0.27 & 1.00 &      \\
$\MTll$ & $166.1\pm\phantom{0}5.7$ & 0.16 & 0.35 & 1.00 \\
\hline
\end{tabular}
\end{center}
\caption[Mtop in each channel]{Summary of the combination of the seven
measurements by CDF and D\O\ in terms of three physical quantities,
the mass of the top quark in the all-jets, lepton+jets, and di-lepton channel. }
\label{tab:three_observables}
\end{table}

\section{Summary}
\label{sec:summary}

A combination of published measurements of the mass of the top quark
from the Tevatron experiments CDF and D\O\ is presented.  The
combination includes five published {\RunI} measurements and two
published {\RunII} measurements.  Taking into account the statistical
and systematic uncertainties and their correlations, the average top
quark mass result is: $\MT= 174.2 \pm 3.3~\GeVc2$.
\vspace*{0.10in}

This average of published results has an accuracy of 1.9\%, while the
present world average top quark mass of $\MT= 172.5 \pm
2.3~\GeVc2$~\cite{Mtop-tevewwgWin06}, which also includes preliminary
analyses of more data by CDF and D\O, has an accuracy of 1.3\%. Both
combinations are limited by the systematic uncertainties, which in
turn are dominated by the jet energy scale uncertainty.  This
systematic is expected to improve as larger data sets are collected
since new analysis techniques constrain the jet energy scale using
in-situ $W\ra qq^{\prime}$ decays. It can be reasonably expected that
with the full \RunII\ data set the top-quark mass could be known to
better than 1\%.  To reach this level of precision further work is
required to determine more accurately the various correlations
present, and to understand more precisely the $b$-jet modeling,
Signal, and Background uncertainties which may limit the sensitivity
at larger data sets.  Limitations of the Monte Carlo generators used
to calibrate each fit method may also become important as the
precision reaches the $1\%$ level and will warrant further study in
the future.

\clearpage

\bibliographystyle{tevewwg}
\bibliography{run2mtop}

\end{document}